\documentclass[aps,pre,reprint]{revtex4-1}

\usepackage{amsmath,amssymb}
\usepackage{graphicx}
\graphicspath{{figures/}}
\usepackage{siunitx}

\renewcommand{\vec}[1]{\mathbf{#1}}
\newcommand{\order}[1]{\mathcal{O}\!\left\{ #1 \right\}}
\newcommand{\abs}[1]{\left\vert #1 \right\vert}

\begin{document}

\title{A perturbative theory for Brownian vortexes}

\author{Henrique W. Moyses}
\author{Ross O. Bauer}
\author{Alexander Y. Grosberg}
\author{David G. Grier}

\affiliation{Department of Physics and Center for Soft Matter
  Research, New York University, New York, NY 10003}

\date{\today}

\begin{abstract}
Brownian vortexes are stochastic machines
that use static non-conservative force fields to bias
random thermal fluctuations into steadily circulating currents
\cite{sun09,sun10}.
The archetype for this class of systems is a colloidal sphere
in an optical tweezer \cite{roichman08a,sun09,demessieres11}.
Trapped near the focus of a strongly converging beam of light,
the particle is displaced by random thermal kicks
into the nonconservative part of the optical force field arising from
radiation pressure \cite{wu09}, which then biases its diffusion
\cite{roichman08a,sun09}.
Assuming the particle remains localized within the trap, its
time-averaged trajectory traces out a toroidal vortex.
Unlike trivial Brownian vortexes, such as the biased Brownian pendulum,
which circulate preferentially in the direction of the bias,
the general Brownian vortex can change direction
and even topology in response to temperature changes.
Here we introduce a theory based on a 
perturbative expansion of
the Fokker-Planck equation for weak non-conservative driving.
The first-order solution takes the form of a modified Boltzmann
relation and accounts for the rich phenomenology
observed in experiments on micrometer-scale colloidal spheres in optical tweezers.
\end{abstract}
\maketitle

\section{Introduction}
\label{sec:introduction}

Stochastic systems can be driven out of equilibrium
in various ways.
Thermal ratchets and Brownian motors use time-dependent forces
to rectify Brownian motion \cite{astumian94,prost94,linke02,astumian02,reimann02}.
Other systems evolve in response to spatial and temporal variations
in the temperature \cite{reimann96,reimann02}.
Even if none of these systems can reach thermodynamic
equilibrium, some can achieve non-equilibrium steady states.
Recently, a distinct class of stochastic machines was
discovered that use noise to extract work
from a static non-conservative force field \cite{roichman08,sun09,simpson10,bammert10}.
These systems, which have been dubbed Brownian vortexes \cite{sun09,sun10},
perform no work at all in the absence of stochastic forces.
When activated by noise, they enter into steady-state motion
characterized by toroidal vortexes in the time-averaged
or ensemble-averaged probability current.
These cyclic processes in principle can be coupled to external systems to extract
useful work.
Since their initial observation in the fluctuations of optically
trapped
colloidal beads \cite{roichman08,sun09},
Brownian vortexes have been reported in trapped colloidal rods
\cite{simpson10},
in trapped spheres subjected to shear flows \cite{bammert10,khan11},
in the response of bacterial populations to antibiotics \cite{bradde12},
and in models for population dynamics \cite{parker11}.

Brownian vortexes arise in time-independent
force fields that include at least one point of stable equilibrium.
In the absence of thermal forces, such systems remain
stationary at their fixed points and so perform no work.
They differ in this respect from conventional machines
that move deterministically under the influence of
non-conservative forces.
Random thermal forces allow a Brownian vortex to explore its force landscape.
Were the force field purely conservative, the system would reach
thermal equilibrium in the Boltzmann distribution, and
so would have no means to perform work.
In force fields with solenoidal components, however, the
probability distribution can be advected by the non-conservative
force.
Probability currents then flow through the system under the
competing influences of advection and diffusion.
These currents must form closed cycles if the system's
overall probability is conserved, raising the possibility that
the system can reach steady state.

Initial studies have identified two broad categories of Brownian vortex
behavior: trivial Brownian vortexes that circulate in the direction
dictated by the non-conservative component of the force landscape,
and general Brownian vortexes that select their own topology
and circulation.
Both types of behavior have been explained with master
equations on discrete networks \cite{sun10}.
Solutions for continuous
systems have been obtained for special cases 
\cite{sun10,parker11,bradde12,russell13}, 
with most examples representing
trivial Brownian vortexes.
Here, we introduce a perturbative theory that accounts
for both trivial and general cases in continuous systems.

\section{Theory}
\label{sec:theory}

Our approach is based on the observation that
systems governed by linear forces
can be mapped to a generalization of the Ornstein-Uhlenbeck process,
for which a general solution is known \cite{kwon05}.
For nonlinear forces, a common approach
is to solve the system using perturbation theory \cite{kwon11}.
Here we develop a perturbation theory for the specific case where 
the non-conservative force is weak compared to the conservative one. 
We show that this approach captures the
important features of general Brownian vortex circulation,
including topological transitions and flux
reversal.   
 
Following earlier studies \cite{sun09,sun10,kwon05,kwon11},
we describe a system capable of undergoing steady-state stochastic circulation
as a Brownian particle of mobility $\mu$ moving through a static
force landscape $\vec{F}(\vec{r})$ at absolute temperature $T$.
Such a description applies naturally to the motion of a colloidal
particle in an optical force field, and also may be applied to more general
systems such as an ensemble of identical particles interacting through
conservative forces and confined by a force landscape.
Assuming that the system reaches steady state, we
seek the steady-state probability distribution $\rho(\vec{r})$
describing the likelihood of finding the particle near position
$\vec{r}$,
and the associated steady-state 
flux of probability $\vec{j}(\vec{r})$ flowing through
that point.
The flux is generated both by advection of the probability
distribution and by diffusion:
\begin{equation}
  \label{eq:flux}
  \vec{j}(\vec{r})
  = 
  \mu \rho(\vec{r}) \vec{F}(\vec{r}) -D \nabla \rho(\vec{r}).
\end{equation}
Because the probability density is non-negative, we
may express it as the exponential of an effective potential
$\phi(\vec{r})$:
\begin{equation}
  \label{eq:density}
  \rho(\vec{r}) = e^{-\beta \phi(\vec{r})},
\end{equation}
so that
\begin{equation}
  \label{eq:flux2}
  \vec{j}(\vec{r})
  =
  \mu \rho(\vec{r}) 
  \left[
    \vec{F}(\vec{r}) + \nabla \phi(\vec{r})
  \right].
\end{equation}

The Helmholtz decomposition theorem
guarantees that the force field  
can be separated into a 
conservative gradient term and a 
non-conservative solenoidal component,
\begin{equation}
  \label{eq:helmholtz}
  \vec{F}(\vec{r}) = -\nabla U(\vec{r}) + \nabla \times \vec{A}(\vec{r}), 
\end{equation}
where $U(\vec{r})$ is the potential energy and $\vec{A}(\vec{r})$ is the  
vector potential.
The probability current may be written in terms of these potentials as
\begin{equation}
  \label{eq:flux2a}
  \vec{j}(\vec{r})
  =
  \mu \rho(\vec{r}) \,
  \left\{
    \nabla \times \vec{A}(\vec{r})
    + 
    \nabla \left[
      \phi(\vec{r}) - U(\vec{r})
    \right]
  \right\}.
\end{equation}

Equation~\eqref{eq:flux2a} makes clear that the system can only
reach thermodynamic equilibrium with $\vec{j}_0(\vec{r}) = 0$ if
the non-conservative force vanishes,
$\nabla \times \vec{A}(\vec{r}) = 0$.
The remaining term in Eq.~\eqref{eq:flux2a} then
yields Boltzmann's distribution for the probability
density,
\begin{equation}
  \label{eq:boltzmann}
  \rho_0(\vec{r}) = e^{- \beta U(\vec{r})},
\end{equation}
with $\phi_0(\vec{r}) = \beta U(\vec{r})$, 
where $\beta^{-1} = k_B T$ is the thermal energy scale.

Systems subject to nonconservative forces
may not reach thermodynamic equilibrium,
but still must satisfy the continuity equation,
\begin{equation}
  \label{eq:continuity}
  \nabla \cdot \vec{j}(\vec{r}) = 
  -\frac{\partial\rho(\vec{r},t)}{\partial t},
\end{equation}
where $\rho(\vec{r},t)$ is the time-dependent probability
density.
Equation~\eqref{eq:continuity} is the Fokker-Planck
equation for Brownian vortexes.
Any steady-state solution of Eq.~\eqref{eq:continuity} satisfies
\begin{equation}
  \label{eq:steadystate}
  \nabla \cdot \vec{j}(\vec{r}) = 0,
\end{equation}
which also ensures conservation of probability.

Quite remarkably, Eq.~\eqref{eq:steadystate}
implies that the steady state distribution $\rho(\vec{r})$ does not depend 
on the particle's mobility, $\mu$, or on any other transport property.
Indeed, substituting Eq.~\eqref{eq:flux2a}
for the current density into Eq.~\eqref{eq:steadystate}
yields an equation that is independent of $\mu$.
This happens because of the assumption, implicit in
Eq.~\eqref{eq:flux}, 
that the particle responds in the same way to an applied force 
whether or not it is conservative.
For a potential force, the Einstein relation 
$D = k_BT\mu$ between the diffusion coefficient $D$ and mobility $\mu$ 
follows from the condition that there exists an equilibrium state 
which obeys the detailed balance condition 
$\vec{j}(\vec{r}) = 0$.
In the presence of a non-conservative force, the system 
does not come to equilibrium, it does not satisfy detailed balance.
Nevertheless, its probability distribution remains independent
of the particle's transport properties, and depends only on the
form of the force field.
In this sense, the steady-state probability distribution resembles
Boltzmann's distribution despite the system's departure from
equilibrium.

To find the specific steady-state probability distribution, we introduce
the projection
\begin{equation}
  \label{eq:projection}
  p(\vec{r}) = \nabla U(\vec{r}) \cdot \nabla \times \vec{A}(\vec{r})
\end{equation}
of the non-conservative force onto the direction of the conservative force.
An exact solution to Eqs.~\eqref{eq:flux2} and \eqref{eq:steadystate}
is known \cite{sun10} only for the special case $p(\vec{r}) = 0$.
In this case, the steady-state probability distribution
retains the form of the Boltzmann distribution, Eq.~\eqref{eq:boltzmann},
and is simply advected by the nonconservative force:
\begin{equation}
  \label{eq:J1}
  \vec{j}(\vec{r}) = \mu e^{-\beta U(\vec{r})} \nabla \times \vec{A}(\vec{r}).
\end{equation}
The direction of $\vec{j}(\vec{r})$ in this case
is fixed by $\nabla \times \vec{A}(\vec{r})$ regardless of the temperature.
Equation~\eqref{eq:J1} therefore describes a trivial Brownian vortex.
Examples of trivial Brownian vortexes include the biased Brownian
pendulum and colloidal spheres circulating in circularly polarized
optical tweezers \cite{ruffner12}.

\begin{figure*}[t]
  \centering
  \includegraphics[width=0.9\textwidth]{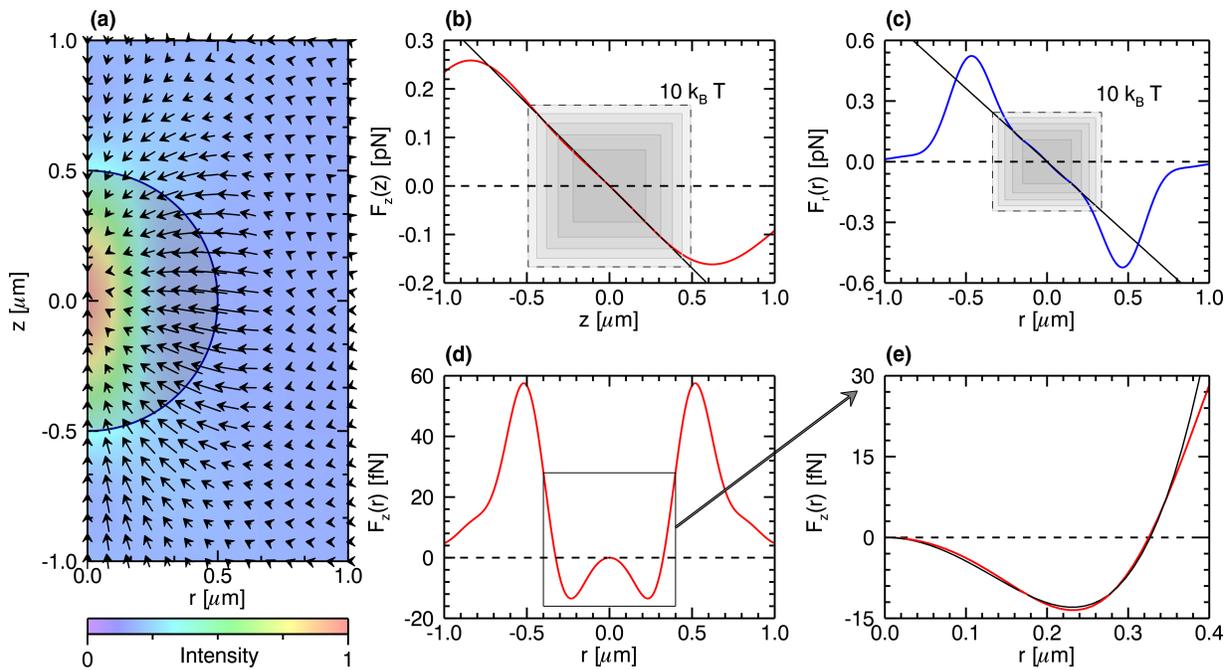}
  \caption{(color online) (a) Force field $\vec{F}(\vec{r})$ 
    experienced by a \SI{1.0}{\um} diameter colloidal silica
    sphere of refractive index 1.45 in a \SI{1}{\milli\watt} Gaussian 
    optical tweezer propagating upward
    along the $\hat{z}$ axis with a \SI{60}{\degree}
    convergence angle.  Colors indicate relative intensity of the
    beam.  Arrows denote direction of net force in the $(r,z)$ plane.
    The superimposed semicircle indicates the size of the colloidal particle.
    (b) Axial component of the restoring force, $F_z(z)$, along $r =
    0$.  
    (c) Radial component of the restoring force, $F_r(r)$, in the
    plane $z = 0$.
    Shaded rectangles in (b) and (c) indicate the particle's range of motion within the force field
    as a function of fluctuation energy in steps of $2~k_B T$ up to
    $10~k_B T$.
    (d) Non-conservative component of the force, $F_z(r)$ in the plane
    $z = 0$.  (e) Detail of (d) showing fit to a quartic polynomial.}
  \label{fig:forcefield}
\end{figure*}

To move beyond this special case, we consider systems in which
$\nabla \times \vec{A}(\vec{r})$ is not necessarily aligned with
$\nabla U(\vec{r})$, but may be treated as a perturbation
whose scale is characterized by a small
parameter $\epsilon$.
We therefore replace $\nabla \times \vec{A}(\vec{r})$ in 
Eq.~\eqref{eq:helmholtz}
with $\epsilon \nabla \times \vec{A}(\vec{r})$.
Assuming the perturbation not to be singular,
the effective potential
may be expanded in orders of $\epsilon$ as
\begin{equation}
  \label{eq:densityexpansion}
  \phi(\vec{r}) 
  =
  \phi_0(\vec{r}) + \epsilon \phi_1(\vec{r}) + \order{\epsilon^2},
\end{equation}
with $\phi_0(\vec{r}) = U(\vec{r})$.
The associated expansion of the probability current,
\begin{equation}
  \label{eq:currentexpansion}
  \vec{j}(\vec{r})
  =
  \epsilon \vec{j}_1(\vec{r}) +
  \order{\epsilon^2},
\end{equation}
has no contribution at zero-th order in $\epsilon$.
The first-order correction,
\begin{subequations}
\label{eq:firstorder}
\begin{equation}
  \label{eq:j1full}
  \vec{j}_1(\vec{r})
  =
  \mu e^{-\beta U(\vec{r})} 
  \left[
    \nabla \times \vec{A}(\vec{r}) +
    \nabla \phi_1(\vec{r}) 
  \right]
\end{equation}
retains the advective term from Eq.~\eqref{eq:J1}, although this now
distorts the probability distribution as well as transporting it.
The second term in Eq.~\eqref{eq:j1full}
describes diffusive relaxation of that distortion.

Because these two terms depend on temperature in different ways
the first-order
expansion admits the possibility of temperature-dependent
transitions such as those characterizing general Brownian vortexes.
These transitions, moreover, may be understood to 
arise from competition between
advection and diffusion.
The balance depends on the distortion of the probability
distribution away from $\rho_0(\vec{r})$.

As explained in Appendix~\ref{sec:perturbationtheory},
substituting Eq.~\eqref{eq:j1full} into
Eq.~\eqref{eq:steadystate}
yields a differential equation for 
$\phi_1(\vec{r})$,
\begin{equation}
  \label{eq:diff_phi1}
  \nabla^2 \phi_1(\vec{r}) - \beta \nabla U(\vec{r}) \cdot \nabla \phi_1(\vec{r})
  =
  \beta p(\vec{r}).
\end{equation} 
Solutions to Eq.~\eqref{eq:diff_phi1} are difficult to find for
arbitrary force fields.
Provided that $U(\vec{r})$ has at least one minimum, however,
the associated operator
$\hat{H}_U = \nabla^2 - \beta \nabla U(\vec{r}) \cdot \nabla$  
has eigenfunctions $\psi_n(\vec{r})$ with eigenvalues $\lambda_n$
that satisfy
\begin{equation}
  \label{eq:eigenfunctions}
  \hat{H}_U \psi_n(\vec{r}) = \lambda_n \psi_n(\vec{r}).
\end{equation}
Symmetrized functions of the form 
$e^{-\frac{1}{2} \beta U(\vec{r})} \psi_n(\vec{r})$
constitute a complete orthogonal basis 
with the orthonormalization condition
\begin{equation}
  \label{eq:orthogonality}
  \int e^{-\beta U(\vec{r})} \psi_n(\vec{r}) \psi_m(\vec{r}) \,
  d\vec{r}
  = \delta_{nm}. 
\end{equation}
We therefore can expand $\phi_1(\vec{r})$ in this basis,
\begin{equation}
  \label{eq:phi1expansion}
  \phi_1(\vec{r}) = \sum_n c_n \psi_n(\vec{r})
\end{equation}
with expansion coefficients
\begin{equation}
  \label{eq:phi1coefficients}
  c_n 
  = 
  \frac{\beta}{\lambda_n} 
  \int e^{-\beta U(\vec{r})}
  \psi_n(\vec{r}) \, p(\vec{r}) \, d\vec{r}.
\end{equation}
\end{subequations}

Equations~\eqref{eq:firstorder} are the principal result of this study.
They describe the lowest-order extension of
a trivial Brownian vortex into a general Brownian vortex under the
influence of a weakly non-conservative force.
The nature of the resulting topological transformations
and flux reversals depends on details of $\vec{F}(\vec{r})$, and
particularly on the projection of its solenoidal component onto its
conservative part.

To illustrate practical applications of Eq.~\eqref{eq:firstorder},
we next apply this formalism to the particular case of colloidal
spheres moving in optical tweezers, the context in which 
Brownian vortexes were first observed \cite{roichman08a,sun09}.
This analysis casts new light on the nature of
this system's behavior.

\section{Colloids in an optical tweezer}
\label{sec:harmonicoscillator}

To facilitate comparisons with experimental studies of colloidal
spheres circulating in optical tweezers, we numerically compute
the forces acting on a trapped sphere with the Lorenz-Mie theory
of light scattering using methods described in
Appendix~\ref{sec:glm}.
Typical results are presented in Fig.~\ref{fig:forcefield}.
Colors in Fig.~\ref{fig:forcefield}(a) represent the relative
intensity of the light in an optical tweezer, as viewed in the
$(r,z)$ plane in cylindrical coordinates.
The light propagates in the $+\hat{z}$ direction (upward)
and comes to a focus along the axis defined by $r = 0$.
Arrows indicate the direction and strength of the resulting force
$\vec{F}(\vec{r})$
experienced
by a colloidal sphere of radius $a_p = \SI{0.5}{\um}$ at position
$\vec{r}$ within that light field.
We have shifted the origin of the coordinate system to coincide
with the position of the trap so that $\vec{F}(0) = 0$.
Away from the origin, $\vec{F}(\vec{r})$ directs the particle
back to the stable fixed point.
All forces reported in Fig.~\ref{fig:forcefield} are computed
for a power of \SI{1}{\milli\watt}, which is a reasonable scale
for typical optical trapping experiments.

Both the axial component of the force, plotted in
Fig.~\ref{fig:forcefield}(b), and the radial component
plotted in Fig.~\ref{fig:forcefield}(c) resemble 
a linear restoring force over a reasonably wide range of axial
displacements.
The axial component of the optical force shows a non-trivial
dependence on radial position, as shown in
Figs.~\ref{fig:forcefield}(d) and (e), that is reasonably
modeled as a quartic polynomial.
This solenoidal dependence is responsible for Brownian vortex
circulation in optically trapped colloidal spheres.

The three-dimensional force field is very nearly symmetric
with respect to rotations about the $\hat{z}$ axis.
Small distortions along the axis of the light's polarization
have little influence on the trapped particle's motions, and
will not be considered here.

\subsection{Simulated circulation}
\label{sec:simulation}

\begin{figure}[!t]
  \centering
  \includegraphics[width=\columnwidth]{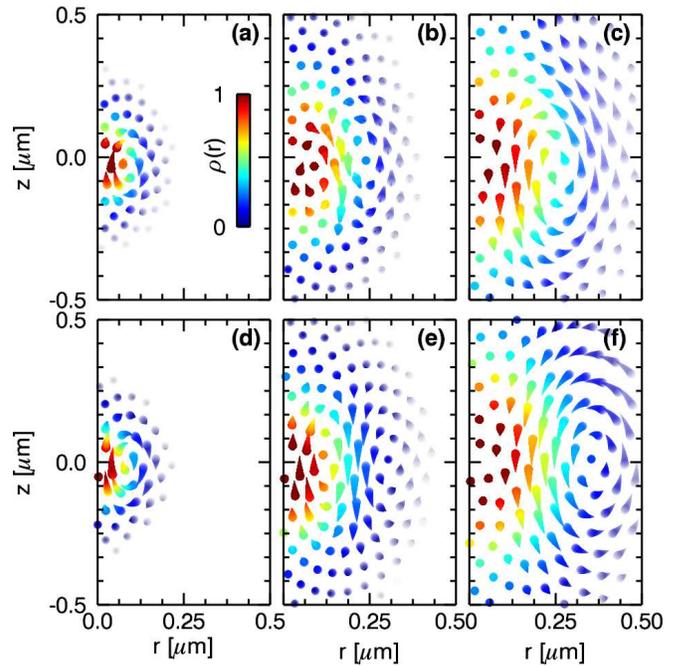}
  \caption{(color online) Streamlines of the current density for a
    \SI{1}{\um}-diameter silica sphere in an optical tweezer as a
    function of trap intensity.  
    Sharp ends of the symbols point in the direction of
    $\vec{j}(\vec{r})$ with colors determined by the relative
    probability $\rho(\vec{r})$, as indicated by the inset color bar.
    Light propagates in the $\hat{z}$ direction.
    (a), (b) and (c) show results from Brownian dynamics simulations in
    force fields computed for a Gaussian optical tweezer at a vacuum
    wavelength of \SI{532}{\nm} and a numerical aperture of 1.4.
    (d), (e) and (f) show corresponding results of
    analytical expressions in Eq.~\eqref{eq:currentdensityresult}
    for the same parameters.
    (a), (d) $\frac{1}{2} \beta k a_p^2 = 30.5$:
    a single toroidal roll is visible near the optical axis.
    (b), (e) $\frac{1}{2} \beta k a_p^2 = 7.6$: concentric
    counter-rotating toroidal vortexes.
    (c), (f) $\frac{1}{2} \beta k a_p^2 = 3.6$:
    weak confinement showing a single flux-reversed roll.
    All of the qualitative features of the Brownian vortex circulation
    observed in the simulation results are obtained also in the
    analytical theory.}
  \label{fig:current}
\end{figure}

The data in Fig.~\ref{fig:current}(a) through \ref{fig:current}(c)
show results of Brownian dynamics simulations
\cite{frenkel01}
of a colloidal silica sphere diffusing through water at $T =
\SI{300}{\kelvin}$
in the computed force field from Fig.~\ref{fig:forcefield}(a).
The sphere's trajectory $\vec{r}_p(t)$ is computed at \SI{100}{\us}
intervals for each of three values of the
light's intensity: \SI{1.38}{\milli\watt} in Fig.~\ref{fig:current}(a),
\SI{0.35}{\milli\watt} in Fig.~\ref{fig:current}(b) and
\SI{0.17}{\milli\watt} in Fig.~\ref{fig:current}(c).
These correspond to values of the radial stiffness, $k$,
of \SI{1}{\pico\newton\per\um}, \SI{0.25}{\pico\newton\per\um}
and \SI{0.12}{\pico\newton\per\um}, respectively.
These values are chosen for consistency with previous experimental studies
\cite{roichman08a,sun09}.
Each trajectory then is compiled into estimates for the
probability density $\rho(\vec{r})$ and the 
probability current density $\vec{j}(\vec{r})$
using an adaptive kernel density estimator
\cite{silverman92,roichman08a,sun09}.
Streamlines of $\vec{j}(\vec{r})$ are denoted by barbs
whose sharp ends point in the direction of motion and whose
length is proportional to the local current density.
Each barb is colored by the estimate for $\rho(\vec{r})$.

At the highest intensity, Fig.~\ref{fig:current}(a),
the trap is stiffest and the particle is most
strongly confined.
A single toroidal roll is evident in the
probability current, and circulates downstream along the optical axis.
Reducing the intensity weakens the trap and 
frees the particle to explore more of the
force landscape.
Under these conditions, shown in Fig.~\ref{fig:current}(b),
streamlines of $\vec{j}(\vec{r})$ reveal
a second counter-rotating roll.
Reference \cite{sun09} identifies the appearance of this second roll
as a topological transition.
Reducing the intensity still further allows the second, outer roll to
dominate the system's dynamics.
It subsumes the inner roll so that only a single toroidal vortex
remains.
The single remaining roll circulates with the flux directed upstream along the optical
axis, which signals a flux reversal relative to Fig.~\ref{fig:current}(a) in addition
to the topological transition relative to Fig.~\ref{fig:current}(b).
Precisely this behavior was reported in \cite{sun09}, which lends
credence both to the experimental results and also to
the present simulations.

The particle's residence time in the trap is effectively indefinite at
the highest laser intensity.  This corresponds to the force field
in Fig.~\ref{fig:forcefield} and the streamlines in
Fig.~\ref{fig:current}(a).
Reducing the laser intensity reduces the residence time, which falls
to \SI{10}{s} under the conditions in Fig.~\ref{fig:current}(c).  This
is consistent with results of experimental studies.
Simulations under these conditions are restarted every time the
particle escapes, and the results averaged to until the computed
current density converges.  The streamlines in
Fig.~\ref{fig:current}(c)
therefore should be viewed as an ensemble average.

\subsection{Analytical model}
\label{sec:model}

The force field $\vec{F}(\vec{r})$ resembles
a cylindrically symmetric harmonic well
for small excursions away from the equilibrium point.
This can be seen in Figs.~\ref{fig:forcefield}(b) and (c).
We therefore model the trapping potential as
\begin{equation}
  \label{eq: potential}
  U(\vec{r}) = 
  \frac{1}{2} k \, (r^2 + \eta \, z^2),
\end{equation}
where $\eta$ characterizes the trap's anisotropy.
In this case, Eq.~\eqref{eq:eigenfunctions}
has a complete set of eigenfunctions,
given in cylindrical coordinates by
\begin{subequations}
\label{eq:harmonicbasis}
\begin{equation}
  \label{eq:ho_eigenfunctions}
  \psi_{\vec{n}}(\vec{r}) =
  N_{\vec{n}} \,
  L_{n}\!\left( \frac{1}{2} \beta k r^2 \right)
  H_{n_z}\!\left( \sqrt{\frac{1}{2} \eta \, \beta k} \, z \right),
\end{equation}
where 
$L_n(\cdot)$ is the Laguerre polynomial of index $n$,
$H_{n_z}(\cdot)$ is the Hermite polynomial of index $n_z$
and where $\vec{n} = \{n, n_z\}$ is a set of whole-number
indexes that are related to the associated eigenvalues by
\begin{equation}
  \label{eq:lambda}
  \lambda_{\vec{n}} = -\beta k ( 2n + \eta \, n_z).
\end{equation}
The harmonic well's basis functions are normalized by
\begin{equation}
  \label{eq:normalization}
  N_{\vec{n}} 
  = 
  \eta^{\frac{1}{4}} \left(\frac{\beta k}{2\pi}\right)^{\frac{3}{4}} 
  \frac{1}{\sqrt{2^{n_z} n_z!}}.
\end{equation}
\end{subequations}
We have chosen basis functions that are independent of $\theta$
to reflect the azimuthal symmetry of $\vec{F}(\vec{r})$.

\subsubsection{Quadratic perturbation in an isotropic trap}
\label{sec:quadratic}

To illustrate applications of Eq.~\eqref{eq:firstorder},
we first consider the symmetric case, $\eta = 1$,
subjected to a quadratic perturbation,
\begin{equation}
  \label{eq:quadraticforce}
  F_z(r) = \epsilon k a_p \left( 1 - \frac{r^2}{a_p^2} \right),
\end{equation}
with $F_x(r) = F_y(r) = 0$,
where $\epsilon$ is the small parameter characterizing the
strength of the non-conservative force and $a_p$ is the
radius of the sphere.
This perturbation is divergence-free and thus has the solenoidal
form assumed in Eq.~\eqref{eq:helmholtz}.
Choosing $F_z(r)$ to be proportional to $k$ ensures that it scales
with the light's intensity in the same way as the linear restoring
force.
Scaling distances by the particle's radius is reasonable because
typical realizations of Brownian vortex circulation involve motions
substantially smaller than $a_p$ \cite{roichman08a,sun09}.
In general, both $k$ and $\epsilon$ depend on the particle's radius relative
to the wavelength of light.
Parameterized in this way, the perturbation may be considered weak if
\begin{equation}
  \label{eq:weakcondition}
  \epsilon < \sqrt{\frac{1}{2} \beta k a_p^2}.
\end{equation}

Choosing a quadratic form for $F_z(r)$ suggests that the non-conservative
force will dominate the linear restoring force at large distances,
and that the particle eventually will escape from the trap.
The rate at which probability leaks from the system is
limited by the smallness of $\epsilon$.
This is the case, for example, in the
system described by Fig.~\ref{fig:forcefield}.
Under these conditions, the particle is likely to remain
trapped over the period of observation, and probability may be
considered to be conserved.
This also is the case for the
experimental studies of optically trapped colloidal particles
\cite{roichman08a,sun09,wu09}
that Eqs.~\eqref{eq: potential} and \eqref{eq:quadraticforce} 
are intended to model.

The projection factor for this system,
defined in Eq.~\eqref{eq:projection},
has the form
\begin{equation}
  \label{eq:quadraticprojection}
  p(\vec{r})
  =
  - k^2a_p z \left(1 - \frac{r^2}{a_p^2} \right).
\end{equation}
It does not explicitly include $\epsilon$ because this
parameter is introduced in Eq.~\eqref{eq:densityexpansion} for
the effective potential and in Eq.~\eqref{eq:currentexpansion} for
the current density.
Substituting this form for $p(\vec{r})$ 
into Eq.~\eqref{eq:firstorder} along with the
basis functions from Eq.~\eqref{eq:harmonicbasis}
yields the first-order correction to the effective
potential,
\begin{equation}
  \label{eq:quadraticpotential}
  \beta \phi_1(\vec{r})
  =
  \frac{1}{3}
  \frac{z}{a_p}
  \left[
    4 - 
    \beta k a_p^2 \left( 3 - \frac{r^2}{a_p^2} \right)\right].
\end{equation}
This correction's linear dependence on $z$ displaces
the probability distribution
down the optical axis, as would be expected of radiation
pressure.
The distribution's width
\begin{equation}
  \label{eq:quadraticwidth}
  \sigma = 
  \left.
    \sqrt{
      \frac{
        \int_0^\infty r^3 \rho(\vec{r}) dr}{
        \int_0^\infty r \rho(\vec{r}) dr}}
  \right\vert_{z=0} 
  = 
  \sqrt{\frac{2}{\beta k}}
\end{equation}
is the same as in the unperturbed case.

The associated current density,
\begin{equation}
  \label{eq:quadraticcurrent}
  \vec{j}(\vec{r}) =
  \frac{2}{3} \epsilon D e^{-\beta U(\vec{r})}
  \left[ \beta k r z \hat{r} + (2 - \beta k r^2) \hat{z} \right],
\end{equation}
has the form of a toroidal roll, centered on the axis $r = 0$
and circulating around a circular core in the plane $z = 0$
at radius
\begin{equation}
  \label{eq:quadraticcore}
  r_0 = \sqrt{\frac{2}{\beta k}},
\end{equation}
which coincides with the distribution's width, $r_0 = \sigma$.
The circulation rate is controlled by the strength $\epsilon$
of the driving force and the rate at which the particle diffuses
in the harmonic potential energy well given its diffusion coefficient
$D = \mu k_B T$.

This Brownian vortex undergoes no topological transitions
or flux reversals.
Applying a quadratic perturbation to a harmonic
well therefore constitutes a model for a trivial Brownian vortex.

\subsubsection{Quartic perturbation in an anisotropic trap}
\label{sec:quartic}

More interesting behavior arises
in a more highly structured force field of the type actually observed in
optical trapping experiments \cite{roichman08a,sun09,wu09}.
Figure~\ref{fig:forcefield}(e)
shows that the computed axial force 
is well approximated near the plane $z = 0$
by a quartic polynomial, 
\begin{equation}
  \label{eq:nconservative2}
  F_z(r)
  = 
  - \epsilon \frac{2 + \eta}{3} k \frac{r^2}{a_p} \left(1 - p_4 r^2 \right),
\end{equation}
with $F_x(r) = F_y(r) = 0$,
where $p_4$ characterizes the particle's interaction
with the focused beam of light.
Like $\epsilon$ and $k$, this additional parameter also depends
on the sphere's radius, $a_p$.
As for the quadratic case, Eq.~\eqref{eq:nconservative2} describes
a purely solenoidal perturbation.
We will show that $p_4$ governs topological transitions and flux
reversals in the resulting Brownian vortex circulation.
Incorporating a constant offset into $F_z(r)$ would move
the equilibrium position along the optical axis as in the quadratic
case, but does not otherwise influence the system's behavior. 
We have omitted such an offset from Eq.~\eqref{eq:nconservative2}
for clarity.
The weak-perturbation condition for this model is
\begin{equation}
  \label{eq:quarticweakcondition}
  \epsilon 
  < 
  \frac{3}{2 + \eta} 
  \sqrt{\frac{1}{2} \beta k a_p^2}.
\end{equation}

Accounting for the trap's anisotropy $\eta$ allows for comparison with
optical trapping experiments.
The force field presented in Fig.~\ref{fig:forcefield}, for example,
is intended to model the influence of Gaussian beam brought
to a diffraction-limited focus and is characterized by $\eta = 0.47$.
Results for isotropic harmonic wells can be retrieved by setting 
$\eta = 1$.  No qualitative features of Brownian vortex circulation
depend on the value of $\eta$.

The projection factor, defined in Eq.~\eqref{eq:projection},
associated with this force field
\begin{equation}
  \label{eq:ho_projection}
  p(\vec{r})
  =
  \frac{2 + \eta}{3} \eta k^2 \frac{z}{a_p} (r^2 - p_4 r^4),
\end{equation}
gives rise to a first-order correction to the effective
potential that also has the form of a quartic polynomial:
\begin{subequations}
\label{eq:densityresult}
\begin{equation}
  \label{eq:densitycorrection}
  \beta \phi_1(\vec{r})
  =
  \frac{2}{3} \frac{z}{a_p} \, \left[
    (1 - 2 \tilde{p}_4) (2 + \tilde{r}^2)
    -
    \frac{2 + \eta}{4 \eta} \, \tilde{p}_4 \tilde{r}^4
  \right].
\end{equation}
For conciseness 
we have introduced the dimensionless parameters
\begin{align}
  & \tilde{r}^2 = \frac{1}{2} \eta \beta k \, r^2 \quad \text{and} \\
  & \tilde{p}_4 = \frac{8}{(4 + \eta) \beta k} \, p_4.
\end{align}
\end{subequations}
As in the quadratic case, the quartic perturbation displaces
probability along $\hat{z}$.
The quartic term also tends to broaden the probability distribution
relative to $\rho_0(\vec{r})$ by introducing a super-Gaussian tail
at large $r$.
This broadening may be of interest for precision measurements
of optical forces because it can influence \cite{roichman08a}
calibration protocols based on analysis of thermal fluctuations
\cite{florin98,polin05,lukic07}.
The availability of an analytical form for $\phi_1(\vec{r})$
will help to assess when nonequilibrium redistribution of the probability
density may be ignored \cite{pesce09}.

The first-order correction to the current density, 
\begin{subequations}
\label{eq:currentdensityresult}
\begin{equation}
  \vec{j}_1(\vec{r}) = j_r(\vec{r}) \hat{r} + j_z(\vec{r}) \hat{z},
\end{equation}
has the radial component
\begin{equation}
  \label{eq:jr}
  \frac{j_r(\vec{r})}{j_0(\vec{r})} =
  \eta \beta k z r
  \left[
    1 - \tilde{p}_4 \left( 2 + \frac{2 + \eta}{2\eta} \tilde{r}^2\right)
  \right]
\end{equation}
that is conveniently measured in units of an overall scale
\begin{equation}
  \label{eq:j0}
  j_0(\vec{r}) 
  = 
  \frac{2}{3} \frac{D}{a_p} e^{-\beta U(\vec{r})}
\end{equation}
that again reflects diffusion in the potential energy well,
as in Eq.~\eqref{eq:quadraticcurrent}.
The radial component of the current density changes sign as it passes through the equatorial plane,
$z = 0$.
The axial component,
\begin{equation}
  \label{eq:jz}
  \frac{j_z(\vec{r})}{j_0(\vec{r})}
  = 
  2 \tilde{p}_4 \frac{2 + \eta}{\eta^2} \, \tilde{r}^4
  -
  4 \left(\frac{1}{\eta} + \tilde{p}_4\right) \, \tilde{r}^2
  +
  4(1 - 2 \tilde{p}_4),
\end{equation}
\end{subequations}
varies non-monotonically with distance $r$ from the axis.
The net current,
$\vec{j}(\vec{r}) = \epsilon \vec{j}_1(\vec{r})$, is proportional
to the strength of the non-conservative driving force.

Streamlines of $\vec{j}(\vec{r})$ plotted in
Fig.~\ref{fig:current}(d) through \ref{fig:current}(f) reveal
circulatory flows that agree well with the simulated results for the
same system under the same conditions.
Specifically, the first-order result for the strongly confined
system in Fig.~\ref{fig:current}(d) shows a single roll consistent
in position, extent and speed with the simulation result in
Fig.~\ref{fig:current}(a).
The double-roll structure in Fig.~\ref{fig:current}(e) similarly
is consistent with that in Fig.~\ref{fig:current}(b), although 
quantitative details of the circulation differ, particularly near
the optical axis.
The single-roll structure in Fig.~\ref{fig:current}(f) again
qualitatively resembles that in Fig.~\ref{fig:current}(c),
although the centers of circulation appear at different
radial positions.
The first-order perturbation theory developed in
Eq.~\eqref{eq:firstorder}
thus captures the essential features
of general Brownian vortex circulation in this model system,
albeit with quantitative discrepancies.
Trends in the analytical results therefore offer useful
insights into the nature of the phenomenon.

\subsection{Topological transition and flux reversal}
\label{sec:transitions}

The probability current, $\vec{j}(\vec{r})$, vanishes at the cores of toroidal
vortexes.
Solutions of $\vec{j}_1(\vec{r}) = 0$ 
take the form of circles in the plane $z = 0$ centered on
the axis at $r = 0$.
Two vortex cores exist if $p_4$ is sufficiently small,
at radii $r_+$ and $r_-$
that satisfy
\begin{subequations}
  \label{eq:vortexcore}
  \begin{align}
    \label{eq:stagnation}
     r_\pm^2 
    & = 
    r_0^2 \, (1 \pm \Delta^2), \quad \text{where} \\
  \label{eq:r0}
    r_0^2 
    & =
      \frac{1}{\beta k} \frac{2}{2 + \eta} \left( \frac{1}{\tilde{p}_4} + \eta \right)
      \quad \text{and} \\
    \label{eq:delta}
    \Delta^2
    & =
      \frac{\sqrt{(8 + 4 \eta + \eta^2) \tilde{p}_4^2 - 4 \tilde{p}_4 +
      1}}{1 + \eta \tilde{p}_4}.
  \end{align}
\end{subequations}

According to Eq.~\eqref{eq:vortexcore},
the vortex at $r_+$ is present for all temperatures greater than zero
and moves outward as the temperature increases.
The other vortex at $r_-$
moves inward, and ceases to exist
when it reaches $r_-=0$, which
occurs when $\tilde{p}_4 = \frac{1}{2}$.
This condition corresponds to a threshold temperature
\begin{equation}
  \label{eq:threshold}
  k_B T_c =\frac{4 + \eta}{16} \, \frac{k}{p_4}
\end{equation}
below which the probability current consists of two counter-rotating
toroidal vortexes and above which only a single roll remains.
The threshold temperature is proportional to the light's intensity through the
trap stiffness, $k$, and depends on details of the particle-light interaction
through $p_4$.
Interestingly, it does not depend on the strength, $\epsilon$, of the
non-conservative force.
In most optical trapping experiments, the temperature
remains constant while $k$ is varied.
Equation~\eqref{eq:threshold} for the threshold temperature
then may be recast into an equivalent condition for the 
trap stiffness.

The system's behavior for $T < T_c$ 
corresponds to strong confinement by
the harmonic well.
The outer toroidal roll therefore 
may not be perceptible in an experiment of finite
duration because the particle spends comparatively little time
exploring the 
outermost reaches of the force landscape.
The apparent transition from one vortex 
in Fig.~\ref{fig:current}(a) to two in Fig.~\ref{fig:current}(b) in
fact results from
increasing the occupation of the outer roll and does not constitute
a topological transition.
The appearance of a second concentric vortex in previous
optical trapping studies \cite{sun09,sun10}
almost certainly corresponds to the same statistical
sampling considerations.

Increasing the temperature beyond $T_c$, or equivalently reducing the stiffness
of the trapping potential, causes an actual topological
transition by eliminating the inner roll at $r_-$.
In this case, the single remaining roll circulates upstream along the
optical axis.
The behavior in Fig.~\ref{fig:current}(f) therefore reflects a flux reversal
relative
to Fig.~\ref{fig:current}(d) as well as a topological transition.
This more dramatic change also has been observed in experimental
studies \cite{sun09}.

\subsection{Size-dependent crossovers}
\label{sec:parameters}

The nature of the Brownian vortex behavior in an optically trapped
colloidal sphere depends qualitatively on the form of the
nonconservative force, $F_z(r)$.
Neither the concentric roll structure nor the high-temperature
topological transition appear if this force is quadratic
rather than quartic.
What dynamic patterns emerge therefore depends
sensitively on the size and composition of the sphere and
the structure of the light.

\begin{figure}[!t]
  \includegraphics[width=\columnwidth]{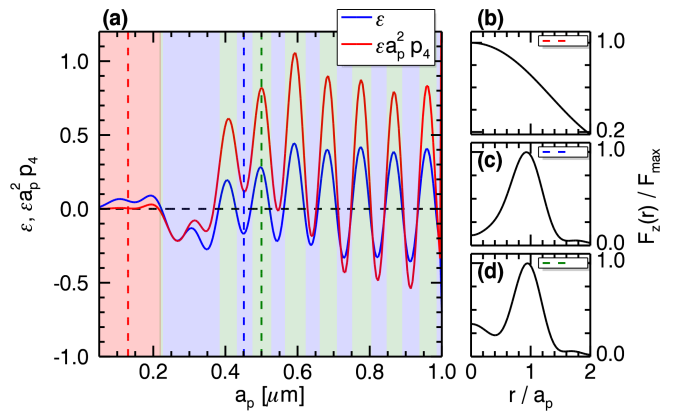}
  \caption{(color online)
    (a) Dependence on the particle radius of the quadratic and quartic coefficients $\epsilon$
    and $p_4$ characterizing the non-conservative optical force
    $F_z(r)$.  These coefficients are obtained by polynomial fits to
    the force field computed with generalized Lorenz-Mie theory
    for a silica sphere in a Gaussian optical tweezer.  Results are
    shown for a silica sphere in water and an optical tweezer created
    with a vacuum wavelength of \SI{532}{\nm} in a beam with a 
    convergence angle of \SI{60}{\degree}.  (b) $F_z(r)$ for a sphere
    with $a_p = \SI{0.13}{\um}$, together with a quartic fit.
    (c) $a_p = \SI{0.45}{\um}$. (d) $a_p = \SI{0.5}{\um}$.
   } 
  \label{fig:fluxreversal}
\end{figure}

The data in Fig.~\ref{fig:fluxreversal} show how the
controlling parameters $\epsilon$ and $p_4$ vary with
sphere radius $a_p$ for a colloidal silica sphere trapped
in water by an optical tweezer of vacuum wavelength \SI{532}{\nm}
brought to a focus by a lens of numerical aperture 1.4.
Figures~\ref{fig:fluxreversal}(b), (c) and (d) show
plots of $F_z(r)$ for the representative radii indicated by
vertical dashed lines in Fig.~\ref{fig:fluxreversal}(a).
As in Fig.~\ref{fig:forcefield}, these results were obtained 
numerically using Eq.~\eqref{eq:maxstreestensor}.

In the Rayleigh limit, for $a_p \ll \lambda$, the non-conservative
force is peaked at the optical axis and falls off 
very nearly quadratically in $r$.
For such particles, $\epsilon > 0$.
Figure~\ref{fig:fluxreversal}(b) is representative of this range 
of conditions.
The quartic contribution being weak,
Rayleigh particles enter into single-roll Brownian vortexes
circulating down the optical axis in the direction of the light's
propagation.
This is the mode of operation that first was identified in
Ref.~\cite{roichman08a}.
The corresponding set conditions is shaded red in
Fig.~\ref{fig:fluxreversal}(a).

Particles that are much larger than the wavelength of light,
$a_p \gg \lambda$,
have force profiles $F_z(r)$ that are peaked far enough from
the optical axis that they also appear to be quadratic
over the accessible range,
but with $\epsilon < 0$.
Such particles circulate in a single toroidal vortex, but in
the sense opposite to that adopted by Rayleigh particles.
The retrograde circulation of larger spheres was
pointed out in a ray-optics analysis of optically trapped colloidal
spheres \cite{huang08}
and subsequently was observed experimentally \cite{sun09}.

Retrograde circulation in a single roll also can arise for particles
that are intermediate in size between the Rayleigh range and
the ray-optics regime.
The example force field depicted in Fig.~\ref{fig:fluxreversal}(c)
has this property, and is characterized by $\epsilon < 0$.  Even though
the quartic term is sizable under these conditions, the curvature of
$F_z(r)$ has the same sign
over the entire accessible range of radii, and only a single 
toroidal roll can be populated.
The domain of such behavior is shaded blue in Fig.~\ref{fig:fluxreversal}(a).

Topological transition and flux reversals are only possible if
$\partial_r F_z(r)$ changes sign in an
accessible part of the force landscape.  That occurs for 
conditions such as those in Fig.~\ref{fig:fluxreversal}(d), and
corresponds to the green-shaded regions in
Fig.~\ref{fig:fluxreversal}(a).
In such cases, the inner roll circulates in the same sense as the
single roll in the Rayleigh regime, and the outer roll rotates in the
opposite sense.
Interestingly, there appears to be no set of conditions that favor
a retrograde double-roll or the corresponding 
topological transition to a retrograde single roll.

Different patterns of size-dependent crossovers 
arise for spheres of different materials, or in beams of different
wavelengths or focusing properties.
Figure~\ref{fig:fluxreversal} makes clear that small variations in
properties can have a large influence on Brownian vortex circulation,
not simply changing the rate of circulation, but rather
reversing the direction of circulation.
The present work has focused on the circulating currents'
topology.  Its results also could be used to address questions
about drift rate and circulation frequency that have been
raised in previous studies \cite{roichman08a,sun09}.

\section{Conclusion}
\label{sec:conclusion}

Brownian vortexes should be generic features of all
probability-conserving stochastic systems subject to
time-independent driving by non-conservative forces.  
Equation~\eqref{eq:firstorder} constitutes 
the leading-order description of a weakly-driven Brownian vortex.
This level of approximation already captures the
topological transitions and flux reversals that have been
reported for archetypal Brownian vortex circulation in optically
trapped colloids.
Analytical results for topological transitions in the current density
go beyond reproducing experimental results by clarifying
their nature and providing a unified
explanation for the system's behavior.
The idealized model of an optical tweezer as a harmonic
well subject to a steady non-conservative force
is likely to serve as a useful model other systems as well.

It is noteworthy that the steady-state
probability distribution in a Brownian vortex,
\begin{equation}
  \label{eq:theanswer}
  \rho(\vec{r}) = e^{-\beta \phi(\vec{r})},
\end{equation}
depends on characteristics of the force field
$\vec{F}(\vec{r})$, but not on
the particle's mobility $\mu$.
The particle's transport properties therefore do not
dictate how the probability redistributes as the system
is driven out of equilibrium.
In this sense, the general relationship between $\vec{F}(\vec{r})$
and the effective potential $\phi(\vec{r})$ 
described in Appendix~\ref{sec:perturbationtheory}
constitute an analog for the Boltzmann relation for
this class of nonequilibrium systems.

The generic nature of the model discussed in
Sec.~\ref{sec:harmonicoscillator}
suggests that this might be a common factor for
steady-state circulation in a broad range of nonequilibrium systems.
The formalism developed here also should apply in other contexts
such as circulatory and oscillatory flows in social networks,
financial systems, and chemical networks.
It also would be interesting to extend this formalism to more
general systems such as systems with multiple fixed
points, and systems of multiple interacting
particles.

\section{Acknowledgments}
\label{sec:acknowledgments}
This work was supported by the National Science Foundation
through Grant Number DMR-135875.

\appendix
\section{Perturbation Theory}
\label{sec:perturbationtheory}

The effective potential, $\phi(\vec{r})$, that determines
the steady-state probability
distribution, $\rho(\vec{r})$,
differs from the imposed potential, $U(\vec{r})$, by an amount
\begin{equation}
  \label{eq:psi}
  \psi(\vec{r}) = \phi(\vec{r}) - U(\vec{r})
\end{equation}
that vanishes if $\nabla \times \vec{A}(\vec{r}) = 0$.
We assume therefore that $\psi(\vec{r})$ is characterized
by the same small parameter, $\epsilon$, that characterizes
$\nabla \times \vec{A}(\vec{r})$.
Imposing conservation of probability through
Eq.~\eqref{eq:steadystate} yields a
differential equation
for $\psi(\vec{r})$,
\begin{multline}
  \label{eq:nonperturbative}
  \left[ 
    \nabla^2 \psi(\vec{r}) 
    - \beta \nabla \psi(\vec{r}) \cdot \nabla U(\vec{r}) 
  \right] \\
  - 
  \left[
    \beta \abs{\nabla \psi(\vec{r})}^2
    + \beta \nabla \psi(\vec{r}) 
    \cdot
    \nabla \times \vec{A}(\vec{r}) \right] \\
  = \beta p(\vec{r}),
\end{multline}
that depends on the temperature and 
characteristics of the force field, but not on the
diffusing particle's mobility.
To first order in $\epsilon$,
this reduces to
\begin{equation}
  \label{eq:firstorderperturbative}
  \nabla^2 \psi(\vec{r}) 
  - \beta \nabla \psi(\vec{r}) \cdot \nabla U(\vec{r}) 
  =
  \beta p(\vec{r}). 
\end{equation}
The associated field
\begin{equation}
  \label{eq:chi}
  \chi(\vec{r}) = e^{-\frac{1}{2} \beta U(\vec{r})} \psi(\vec{r})
\end{equation}
then may be obtained as an expansion in eigenfunctions
$\chi_n(\vec{r})$ of the Hermitian operator
\begin{equation}
  \label{eq:hermitianoperator}
  \hat{H}^\prime_U 
  = 
  \nabla^2 
  + \left[
    \frac{1}{2} \beta \nabla^2 U(\vec{r})
    - \frac{1}{4} \beta^2 \abs{\nabla U(\vec{r})}^2 
  \right].
\end{equation}
Specifically, solutions of
\begin{equation}
  \label{eq:eigenequation}
  \hat{H}^\prime_U \chi_n(\vec{r}) = \lambda_n \chi_n(\vec{r})
\end{equation}
form a complete set of basis functions labeled by index $n$
with eigenvalues $\lambda_n$.
When appropriately normalized they
satisfy the orthogonality condition
\begin{equation}
  \label{eq:chiorthogonality}
  \int \chi_m(\vec{r}) \chi_n(\vec{r}) \, d^3r = \delta_{mn}.
\end{equation}
Equation~\eqref{eq:orthogonality} follows from
Eq.~\eqref{eq:chiorthogonality}.

Imposing conservation of probability at each order of the perturbation
expansion ensures that solutions reflect steady state behavior.
Higher-order corrections obtained by incorporating the second-order
terms in Eq.~\eqref{eq:nonperturbative} will redistribute the
probability distribution, but are not likely to eliminate
qualitative features of the steady-state circulation that arise at
first order.

\section{Generalized Lorenz-Mie theory}
\label{sec:glm}

An optical tweezer can be modeled as a strongly focused Gaussian beam,
and its field can be expanded as a series in vector spherical
harmonics,
\begin{equation}
  \label{eq:incexpansion}
  \vec{E}_i(\vec{r}) = E_0 
  \sum_{n=1}^{\infty}  \sum_{m=-n}^{n}
  \left[a_{mn} \vec{M}^{(1)}_{mn}(k\vec{r}) +
          b_{mn} \vec{N}^{(1)}_{mn}(k\vec{r}) 
          \right].
\end{equation}
The coefficients $a_{mn}$ and
$b_{mn}$ have been previously reported for a converging Gaussian beam
\cite{bohren83,mishchenko02,taylor09}. 
and depend on the numerical aperture of the lens that
brings the beam to a focus.

The incident beam illuminates a particle located at $\vec{r}_p$, which
gives rise to a scattered wave $\vec{E}_s(\vec{r} - \vec{r}_p)$ that
propagates to position $\vec{r}$.
The corresponding expansion for the scattered field,
\begin{equation}
  \label{eq:sctexpansion}
  \vec{E}_s(\vec{r}) = E_0 
  \sum_{n=1}^{\infty}  \sum_{m=-n}^{n}
  \left[r_{mn} \vec{M}^{(3)}_{mn}(k\vec{r}) +
          s_{mn} \vec{N}^{(3)}_{mn}(k\vec{r}) 
          \right],
\end{equation}
has expansion coefficients,
\begin{equation}
  \label{eq:sctexpansioncoeff}
  \begin{split}
  r_{mn} &= -b_n \ a_{mn} \ \ \text{and} \\
  s_{mn} &= -a_n \ b_{mn},
  \end{split}
 \end{equation}
that are related to the incident beam's coefficients by the
particle's Lorenz-Mie scattering coefficients,
$a_n$ and $b_n$ \cite{bohren83,mishchenko02}.
For the particular case of scattering by a sphere, the 
Lorenz-Mie coefficients depend on the sphere's 
radius and refractive index relative to the medium.

The combined incident and scattered fields, 
\begin{equation}
  \label{eq:totalfield}
  \vec{E}(\vec{r}) 
  = 
  \vec{E}_i(\vec{r}) + \vec{E}_s(\vec{r} - \vec{r}_p),
\end{equation}
can be used
to calculate the Maxwell stress tensor, $\vec{T}(\vec{r})$,
with components
\begin{equation}
  \label{eq:maxwellstresstensor}
  T_{ij}(\vec{r}) 
  = 
  \epsilon_m E_i(\vec{r}) E_j(\vec{r}) -
  \frac{1}{2} \epsilon_m E^2(\vec{r}) \, \delta_{ij},
\end{equation}
where $\epsilon_m$ is the dielectric constant of the medium.
The optically induced force
is then obtained by integrating the Maxwell stress tensor
over a surface $S$
that encloses the sphere,
\begin{equation}
  \label{eq:maxstreestensor}
  \vec{F(\vec{r})} = 
  \oint_S \hat n \cdot
  \vec{T}(\vec{r^\prime}) \, d\vec{r^\prime},
   \end{equation}
where $\hat{n}$ is the unit vector normal to $S$.
Techniques to
perform this integration have been previously reported
\cite{alexander89,nieminen07,sun08,ruffner14}.

Figure~\ref{fig:forcefield}
shows the computed force field acting on
a \SI{0.5}{\um}-radius silica sphere in water in
an optical tweezer
with vacuum wavelength  $\lambda = \SI{532}{\nm}$
projected by a lens with numerical aperture 1.4 (convergence
angle \SI{60}{\degree}).
We define the particle's point of mechanical equilibrium
to be
the origin of the coordinate system in Fig.~\ref{fig:forcefield}.

%

\end{document}